\documentclass[a4paper,fleqn]{cas-dc}
\usepackage{cite}
\usepackage[square, comma, sort&compress, numbers]{natbib}
\usepackage{enumitem}
\setenumerate[1]{itemsep=0pt,partopsep=0pt,parsep=\parskip,topsep=5pt}
\setitemize[1]{itemsep=0pt,partopsep=0pt,parsep=\parskip,topsep=5pt}
\setdescription{itemsep=0pt,partopsep=0pt,parsep=\parskip,topsep=5pt}
\usepackage{subfigure}
\usepackage{flushend}
\usepackage{amsmath}
\usepackage{amssymb}
\usepackage{graphicx}
\usepackage{url}

\begin{document}
\let\printorcid\relax
\let\WriteBookmarks\relax
\def\floatpagepagefraction{1}
\def\textpagefraction{.001}

\title[mode = title]{Online Network Traffic Classification Based on External Attention and Convolution by IP Packet Header}

\tnotemark[1]

\author[author1]{Yahui Hu}
\author[author1]{Ziqian Zeng}
\cormark[1]
\cortext[cor1]{Corresponding author}
\author[author2]{Junping Song}
\cormark[1]
\author[author2]{Luyang Xu}
\author[author2]{Xu Zhou}
\address[author1]{China University of Mining and Technology (Beijing), Beijing 100083, China}
\address[author2]{Computer Network Information Center, Chinese Academy of Sciences, Beijing 100190, China}

\tnotemark[1]

\tnotetext[1]{This work was supported by the National Natural Science Foundation 
  of China (Grant No.U1909204), the Youth Innovation Promotion Association
  of Chinese Academy of Sciences (2021168).}

\begin{abstract}[S U M M A R Y]
	Network traffic classification is an important part of network monitoring and network management. Three traditional methods for network traffic classification are flow-based, session-based, and packet-based, while flow-based and session-based methods cannot meet the real-time requirements and existing packet-based methods will violate user's privacy. To solve the above problems, we propose a network traffic classification method only by the IP packet header, which satisfies the requirements of both the user's privacy protection and online classification performances. Through statistical analyses, we find that IP packet header information is effective on the network traffic classification tasks and this conclusion is also demonstrated by experiments. Furthermore, we propose a novel external attention and convolution mixed (ECM) model for online network traffic classification. This model adopts both low-computational complexity external attention and convolution to respectively extract the byte-level and packet-level characteristics for traffic classification. Therefore, it can achieve high classification accuracy and low time consumption. The experiments show that ECM can achieve the highest classification accuracy and the lowest delay, compared with other state-of-art models. The accuracy can respectively achieve 98.39\% and 95.57\% on two datasets and the classification time is shorten to meet the real-time requirements.
\end{abstract}

	
		

\maketitle

\section{Introduction}\label{Indro}
Network traffic classification is of great significance to network operation management. How to improve the classification accuracy and real-time performance has been always a key issue in the field of network traffic classification. At the same time, with the excellent performance of deep learning in machine vision and other fields, its application to network traffic classification has become a hot research topic in recent years \cite{gu2021research,rezaei2019deep,wang2019survey}.

From the perspective of the granularity of the network traffic data to be mined, there are mainly three kinds of related works: 1) the packet-level traffic identification \cite{lotfollahi2020deep,zeng2019test,xu2019traffic,ren2021tree,xie2021self}, i.e., the traffic category of each packet in the network traffic is identified separately; 2) the flow-level traffic identification \cite{wang2017end,liu2019fs,shapira2021flowpic,zhao2021flow,babaria2021flowformers}, i.e., identifying the upstream or downstream traffic in the typical "client-server" or "browser-server" model communication process; 3) the session-level traffic identification \cite{wang2017end,yang2021aefeta}, i.e., the entire upstream and downstream traffic of the client or browser and server communication is considered as a whole, and the identification is conducted based on the session. Among them, session-level and flow-level identifications require waiting for sufficient traffic packets to obtain meaningful time series and statistical features. They cannot meet the high real-time requirements of the online network traffic classification tasks \cite{gu2021research, wu2022online}. In contrast, each packet's waiting time is much shorter, and thus, packet-level identifications are more suitable for online traffic classification. 

Unfortunately, most of the current packet-level online traffic classification methods need to exploit payload data and inevitably violate the user's privacy protection requirements \cite{gu2021research,lotfollahi2020deep,li2018byte,zeng2019test,xu2019traffic,ren2021tree}. Therefore, it is necessary to study whether network traffic can be classified only using packet headers instead of data payloads in order to meet the two-tier requirements of classification performance and user's privacy protection. So far as we know, the shortest data packet length used for network traffic classification is 40 bytes in all the previous works \cite{xie2020sam,xie2021self}. It actually does not involve payload data when the application transmission protocol is TCP/IP. However, it does not work when the service transmission protocol is UDP/IP since the packet header length is less than 40 bytes, and the payload data is inevitably used as one part of raw data. In summary, no research work is clearly proposed to devote to the problem of network traffic classification with only packet headers. In general, it may be believed that IP packet headers do not carry service content information and the feasibility of classifying network traffic only by packet header is doubtful. To solve this problem, we propose the idea of network traffic classification only based on the IP packet header and do some works to demonstrate its feasibility in this paper.

Firstly, we conduct statistical analyses on the IP packet header information. Two fields of source and destination IP addresses are eliminated, which have little relation with the traffic types. The other remaining fields are constituted by 12 bytes and the value of each byte is treated as a random variable. The joint distribution of these random variables are studied by statistics and the statistical characteristics of some bytes differ among different network traffic types, more details about the statistical outcomes can be found in section 3. The results show that IP packet header can be used as traffic characteristics for classification. However, it should not be ignored that the service type information contained in the 12 bytes of the IP packet header is far less rich than that contained in the 40/50 bytes composed of the packet header and payload. Therefore, how to effectively mine IP header information without reducing classification accuracy and meet the low-delay requirements of online classification is a major challenge. To address this challenge, we design a novel external attention and convolution mixed (ECM) model to further improve the classification accuracy and model inference speed, which are the most critical performance indicators for classification tasks.

The proposed ECM model consists of five modules, i.e., data preprocessing, the embedding layer, the external attention layer, the convolutional layer, and the linear layer. Data processing and embedding layer are responsible for converting the raw IP packet files to high-dimensional feature data of IP packet header. Then, the feature data is put into the learning part, including external attention layer, convolution layer and linear layer. The external attention layer is applied to deal with the intra-byte data and extracting helpful byte-level information for traffic classification. Compared with self-attention, the external attention mechanism can achieve better performance with lower computational complexity for visual tasks \cite{guo2021beyond}. In this paper, we exploit this merit to improve the classification accuracy and speed. Furthermore, the convolution layer is designed to capture the inter-byte information and obtain packet-level semantics, which can further improve the classification accuracy. Finally, the linear layer makes the decision for the corresponding network traffic category.

In summary, the main contributions of this paper are as follows.
\begin{itemize} 
              \item We conduct a statistical analyses of the byte information within the IP packet headers. The outcomes show that distributions of the byte information are actually different among network traffic. Therefore, using only IP packet headers for network traffic classification is proposed to meet online classification requirements and avoid the privacy problems caused by using payload information.
              \item An external attention and convolution mixed (ECM) model for network traffic classification is proposed. It sufficiently extracts semantic information at both the byte-level and packet-level from the 12-byte input length of IP packet header through external attention and CNN respectively.
              \item Experiments are made on two different datasets, i.e. the public ISCX and our private dataset. The average accuracy is respectively 98.39\% and 95.57\%, and the classification time is about 0.36ms per packet. Furthermore, baselines comparison experiment shows the ECM model almost outperforms state-of-art solutions both on the classification accuracy and real-time performance. 
\end{itemize}

The rest of this paper is organized as follows. We review recent network traffic classification studies in Section 2. In Section 3, we conduct a statistical analysis of the value distribution within the IP packet header. Section 4 illustrates the details of the model we proposed. We discuss the results of our experiments in Section 5, and finally, in Section 6, we summarize and prospect our work.

\section{Related Work}
Compared with machine learning algorithms, deep learning algorithms have the advantage of exploiting the deeper and higher semantic features of the input network traffic data and are more suitable for diverse network traffic classification scenarios. Especially in recent years, with the successful application of deep learning models in natural language processing and computer vision, researchers have also applied them to network traffic classification scenarios. Approaches based on deep learning can be further divided into session-based, flow-based, and packet-based ones.

The input data of session-based and flow-based deep learning approaches are usually statistical features or raw byte data. Wang et al. \cite{wang2017end} utilize the first 784 bytes of a service flow and transform the byte data into a square grayscale image. Then, a one-dimensional convolutional neural network for network traffic classification is adopted to achieve over 90\% accuracy on the UNB ISCX VPN-nonVPN dataset. Based on the sequence features of packets, Liu et al. \cite{liu2019fs} combine recurrent neural networks with auto-encoder, eventually achieving an average classification accuracy of 99\% on 18 services, such as Alipay, QQ, Weibo, and Taobao. Two features of a flow, i.e., packet arrival interval time and packet size, are constructed into the FlowPic, which is the input of LeNet-5 neural network \cite{shapira2021flowpic}. And it achieves an average accuracy of 93\% and 97\% on Non-VPN and VPN flows. The first 4 packets of each session and the first 120 bytes of each packet are used as the input data, and the combined model of CNN, LSTM, and attention mechanism are proposed in \cite{yang2021aefeta} to automatically extract spatio-temporal features to perform the classification task. The recall and accuracy performance are respectively 98\% and 97\% on six classes of encrypted traffic from the UNB ISCX VPN-nonVPN dataset. Zhao et al. \cite{zhao2021flow} apply random forest algorithm to select 84 important statistical features of a flow and then use the Transformer model to classify each flow, which can achieve 86\% and 95.2\% accuracy on ten services of SJTU-AN21 and seven services of UNB ISCX VPN-nonVPN. In \cite{babaria2021flowformers}, the number of packets of bidirectional data flow and the byte size of the packets are exploited to construct the flow’s "fingerprint information" and put the fingerprint information into the Transformer model for classification, which respectively achieves 97\% and 95\% F1-score for video-based services (Netflix, Youtube and five other categories) and meeting applications (Zoom, Microsoft Teams and five other categories). Lin et al. \cite{lin2022efficient} propose a network structure that combines CNN and Bi-GRU and selects the first five packets of a session and the first 512 bytes of each packet as input. Finally, the model has a classification accuracy of 93.1\%, with a recall rate of 93.7\% and an F1-score of 93.6\% on the public dataset UNB ISCX VPN-nonVPN. Wang et al. \cite{wang2022sessionvideo} propose a deep learning method based on 3D-CNN to capture the temporal characteristics of network traffic. By using the first eight packets of a session and the first 1024 bytes to construct "SessionVideo", they can achieve an accuracy of 97.89\% and a weighted average precision of 97.96\% on a traffic dataset of 20 applications. Yang et al. \cite{yang2023network} use both packets' length sequence and packet bytes as model input and diverge into two paths to analyze the dual-model features of hybrid neural networks. The experiments show that the accuracy of this method exceeded 91\% across four datasets. The above existing session-based and flow-based classification methods obtain outstanding classification performance in terms of accuracy. However, the time consumption of obtaining a sufficient number of packets in a flow/session and collecting statistical information about these packets is usually too high to meet the needs of online traffic classification. In paper \cite{wu2022online}, the author calculates the execution time of different methods, including FE (Feature Extraction) time and inference time. It can be seen that flow-level or session-level based methods consume a higher amount of FE time, ranging from a few milliseconds or tens of milliseconds to over 100 milliseconds. Therefore, some researchers focus on directly exploiting the raw byte data of each individual packet.

\begin{table*}[htbp,h]
	\normalsize
	\centering
	\caption{Summary of network traffic classification using deep learning methods}
	\scalebox{0.5}{
	  \begin{tabular}{cccccc}
	  \toprule
	  Paper & TC object & Input Data & Classifier & Experimental Results & Year \\
	  \midrule
	  Wang et al.\cite{wang2017end} & Flow/Session & ALL/L7 Layers[784 B] & 1D-CNN & 86.6\% acc.(12 classes) & 2017 \\
	  \midrule
	  Liu et al.\cite{liu2019fs} & Flow  & Packet Length Sequences & GRU   & 99.14\% TPR,0.05\% FPR,0.9906 FTF(18 apps) & 2019 \\
	  \midrule
	  Shapira et al.\cite{shapira2021flowpic} & Flow  & IP Packet Size\&Time of Arrival & LeNet-5 & 93.76\%/97.59\% acc.(Non-VPN/VPN 5 classes) & 2021 \\
	  \midrule
	  Yang et all.\cite{yang2021aefeta} & Session & 4 Packets[120 B]in per Session & CNN+BiLSTM+Attention & 98\% rec. \& 97\% pre.(6 classes) & 2021 \\
	  \midrule
	  Zhao et all.\cite{zhao2021flow} & Flow  & 84 Statistical Features & Transfomer & 86\%/95.2\% acc. (10 classes/7 classes) & 2021 \\
	  \midrule
	  Babaria et all.\cite{babaria2021flowformers} & Flow  & Up/Down Packets/Bytes & Transfomer & 97\% f1.(5 apps),95\% f1.(4 videos/4 conferences)  & 2021 \\
	  \midrule
	  Lin et all.\cite{lin2022efficient} & Session & 5 Packets[512 B]in per session & CNN+Bi-GRU & 93.1\% acc. 93.7\% rec. 93.6\& f1.(12 classes) & 2022 \\
	  \midrule
	  Wang et all.\cite{wang2022sessionvideo} & Session & 8 Packets[1024 B]in per session& 3D-CNN & 97.89\% acc.(20 apps) & 2022 \\
	  \midrule
      Yang et all.\cite{yang2023network} & Flow & packets' length sequence+packets bytes & GRU+SAE & more than 91\% acc. on four datasets & 2023 \\
	  \midrule
	  Lotfollahi et al.\cite{lotfollahi2020deep} & Packet & Header+Payload[1500 B] & SAE,1D-CNN & 98\%/94\% recall.(17 / 12 classes) & 2019 \\
	  \midrule
	  Zeng et all.\cite{zeng2019test} & Packet & Header+Payload[900 B] & CNN+LSTM & 99.98\% acc.(8 classes) & 2019 \\
	  \midrule
	  Xu et all.\cite{xu2019traffic} & Packet & Header+Payload[784 B] & 1D-CNN,2D-CNN,LSTM & 96.38 acc.(6 classes) & 2019 \\
	  \midrule
	  Ren et all.\cite{ren2021tree} & Packet & Header+Payload[784 B] & TREE-RNN & 98.98 acc.(6 classes) & 2021 \\
	  \midrule
	  Xie et all.\cite{xie2021self} & Packet & Header+Payload[50 B] & Attention+CNN & 94\%/92\%/90\% acc.(6 types of traffic for protocol/application/traffic type classification) & 2021 \\
	  \bottomrule
	  \end{tabular}%
	}
	\label{tab:1}%
  \end{table*}%
  
Lotfollahi et al. \cite{lotfollahi2020deep} propose to extract the first 1500 bytes of a packet and adopt a combined model of a one-dimensional convolutional neural network (1D-CNN) and a stacked auto-encoder (SAE) for a classification task. The average recalls are 98\% on a 17-class coarse-grained application classification task and 94\% on a 12-class fine-grained service classification task. In \cite{li2018byte}, packets are firstly split into segments of every eight bytes. The encoded segments are input into the Gated Recurrent Unit (GRU) and attention models for protocol classification. Experiments demonstrate the average F1-score is about 95.82\% on five protocols, i.e., QQ and the other four protocols. Zeng et al. \cite{zeng2019test} propose to use the first 900 bytes of a packet and a combined CNN and LSTM structure to extract service-related temporal and spatial features, and it is able to achieve 99.98\% accuracy on the combined dataset of UNB ISCX VPN-nonVPN and USTC-TFC2016. Xu et al. \cite{xu2019traffic} choose to make use of the first 784 bytes of a single packet and integrated learning models to elect the classification result from the outcomes of one-dimensional CNN, two-dimensional CNN, and LSTM classifiers. Finally, it achieves over 96\% accuracy on the UNB ISCX VPN-nonVPN dataset. Ren et al. \cite{ren2021tree} also use the first 784 bytes of a packet as the classification raw data and propose a TREE-RNN model for classification, achieving 98\% accuracy on the UNB ISCX VPN-nonVPN dataset. Furthermore, it is claimed as an online approach for its high classifying speed of 5.4 packets per ms.

Although the above packet-based classification methods can achieve relatively satisfactory accuracy, the amount of data used is usually up to hundreds or thousands of bytes, which involves the load part of the application layer and cannot meet the user privacy protection requirements. Because of this, the model SAM built by Xie et al. \cite{xie2020sam,xie2021self} using the self-attention mechanism, which, unlike other researchers, uses the first 50 bytes of a packet and takes the raw bytes of the packet header as model input has a high F1-score on the UNB ISCX VPN-nonVPN dataset for protocol classification (98.62\%) and application classification (98.93\%). For the method employing the first 50 bytes of packet header, network traffic packets using TCP protocol do not involve application layer data. However, this method still violates users' information data for UDP network traffic packets.

It is summarized in Table 1 that the above classification methods, the comparative analyses in terms of TC object, input data, classifier, experimental results, and years of publication, where TC denotes Traffic Classification, acc. denotes accuracy, rec. denotes recall and f1. denotes f1-score. From Table 1, it is obvious that almost all existing network traffic classification methods involved the payload, and therefore, they cannot meet the demand of protecting user privacy. Therefore, it is necessary to explore the possibility of classifying network traffic only on the IP packet header. Another conclusion is that the deep learning models used for network traffic classification have been optimized with the self-development of deep learning, and the latest results are mainly self-attention-based models. Therefore, it is also necessary to design an effective network traffic classification model based on the latest deep learning technologies to further improve the online classification performance, such as accuracy and speed.

\section{Statistical Feature Analyses Of IP Packet Header}

\begin{figure}[htbp]
	\centerline{\includegraphics[scale=0.5]{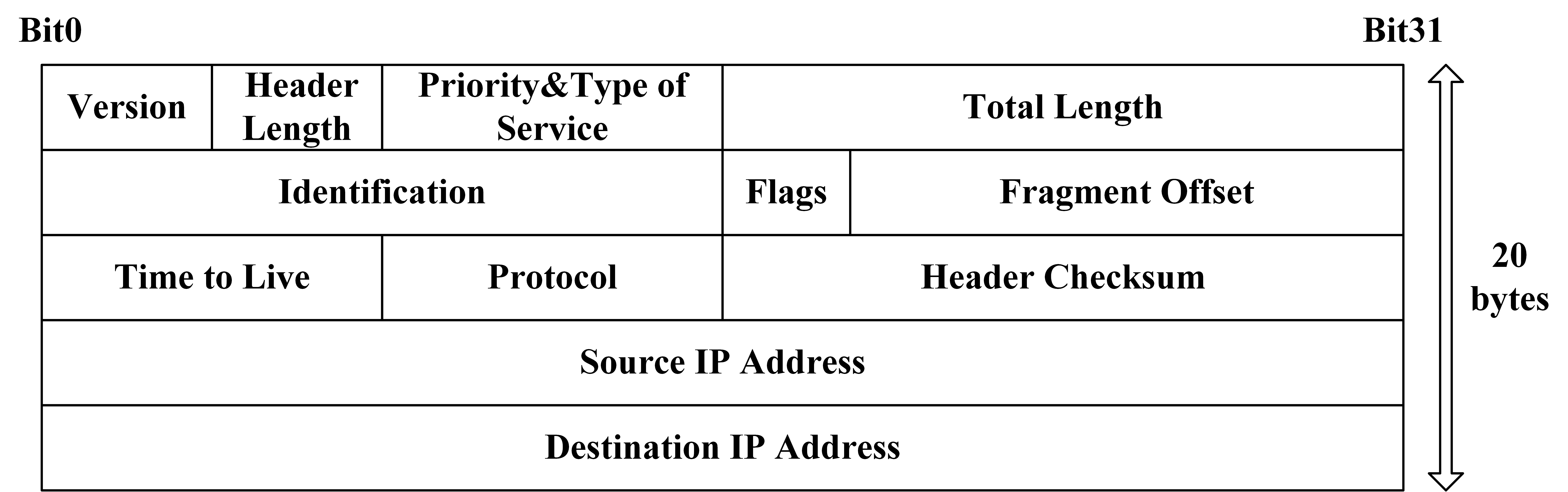}}
	\caption{IPV4 Network Layer Header Structure} 
	\label{fig:1} 
\end{figure}

\begin{figure*}[H]
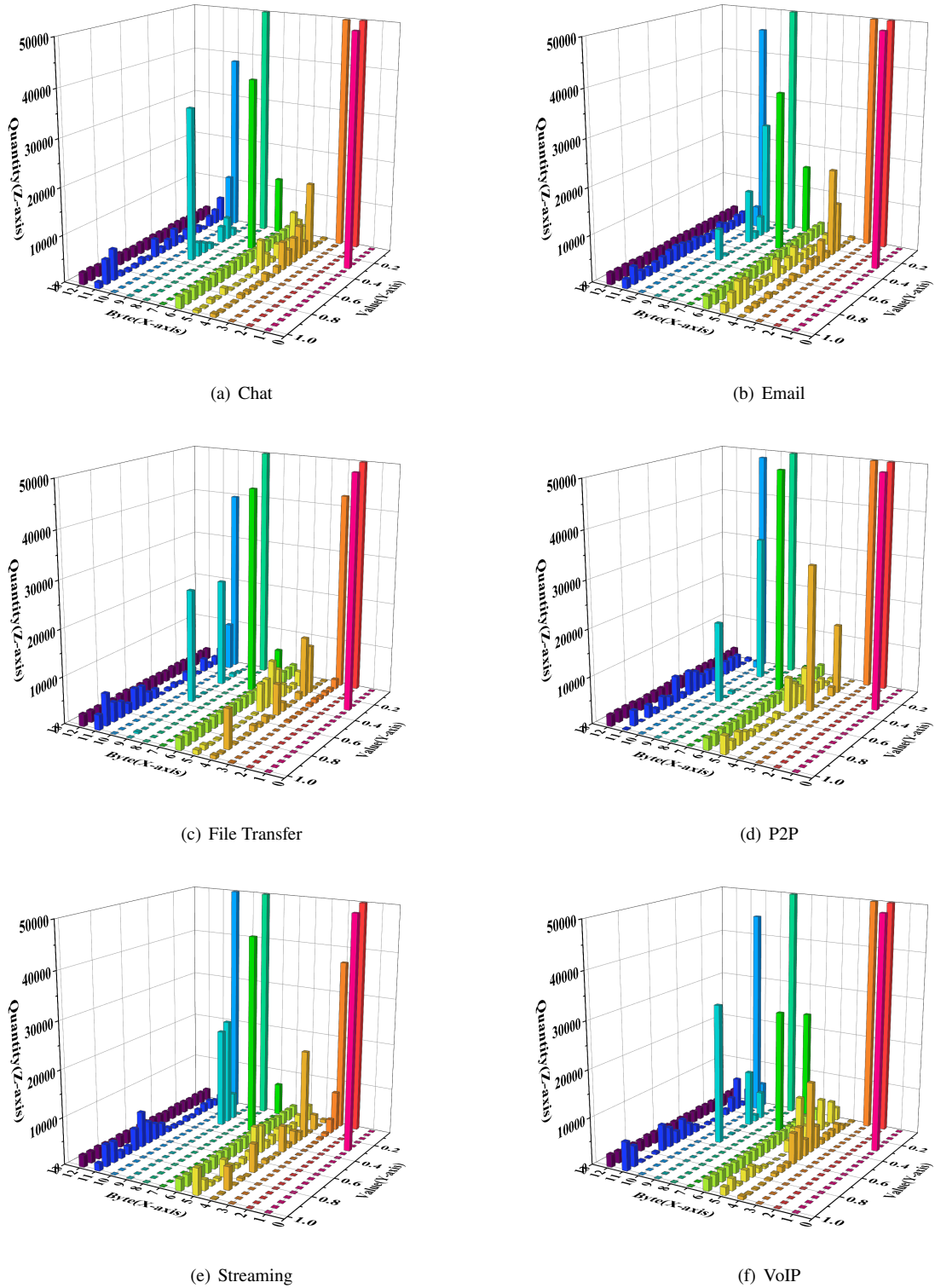

	\centering
	\subfigure[Chat]{
		\label{fig:subfig_a}
		\includegraphics[width=0.45\textwidth]{Chat.png}
	}
	\subfigure[Email]{
		\label{fig:subfig_b}
		\includegraphics[width=0.45\textwidth]{Email.png}
	}
	\\
	\subfigure[File Transfer]{
		\label{fig:subfig_c}
		\includegraphics[width=0.45\textwidth]{FileTransfer.png}
	}
	\subfigure[P2P]{
		\label{fig:subfig_d}
		\includegraphics[width=0.45\textwidth]{P2P.png}
	}
	\\
	\subfigure[Streaming]{
		\label{fig:subfig_e}
		\includegraphics[width=0.45\textwidth]{Streaming.png}
	}
	\subfigure[VoIP]{
		\label{fig:subfig_f}
		\includegraphics[width=0.45\textwidth]{VoIP.png}
	}
	\caption{The statistical distribution of IP packet header byte information on ISCX dataset.}
\end{figure*}

The network layer packet header in the IPV4 network generally contains a fixed length of 20 bytes, whose structure is shown in Fig. 1. And the IP Packet header is further divided into the following 12 fields, i.e., Version, Header Length, Priority\&Type of Service, Total Length, Identification, Flags, Fragment Offset, Time to Live, Protocol, Header Checksum, Source IP Address and Destination IP Address. In particular, some fields, such as source/destination IP address, are closely related to the local network configuration and have little relation with the traffic types \cite{xie2021self,aceto2020toward,aceto2021distiller}. Therefore, they are redundant information and would affect the generalization ability of the model and the results of classification from the viewpoint of network traffic classification task. As a result, in many traffic classification schemes, the source/destination IP address are usually masked with zeros in the raw data preprocessing. 

In our scheme, we choose to directly remove the 8-byte length source and destination IP addresses from the network layer IP packet header, further reducing the amount of data input into the neural network to shorten the inference time. For the remaining 12 bytes, they correspond to different fields within the IP packet header that serve various roles and functionalities: Byte 1 represents Version and Header Length, Byte 2 represents Priority\&Type of Service, Bytes 3 and 4 represent Total Length, Bytes 5 and 6 represent Identification, Byte 7 represents a portion of Flags and Fragment Offset, Byte 8 represents the remaining portion of Fragment Offset, Byte 9 represents Time to Live, Byte 10 represents Protocol, and Bytes 11 and 12 represent Header Checksum.

\begin{table}[htbp,h]
	\normalsize
	\centering
	\caption{ISCX Dataset categories and numbers of samples}
	\scalebox{0.7}{
		\begin{tabular}{ccc}
			\toprule
			Traffic Type & Content & Quantity \\
			\midrule
			\multirow{5}[2]{*}{Chat} & AIM\_Chat & \multirow{5}[2]{*}{50000} \\
				  & facebook\_chat &  \\
				  & hangouts\_chat &  \\
				  & ICQ\_chat &  \\
				  & Skype\_Chat &  \\
			\midrule
			\multirow{2}[2]{*}{Email} & Email & \multirow{2}[2]{*}{50000} \\
				  & Gmail &  \\
			\midrule
			\multirow{4}[2]{*}{File Transfer} & FTPS  & \multirow{4}[2]{*}{50000} \\
				  & SCP   &  \\
				  & SFTP  &  \\
				  & Skype\_File &  \\
			\midrule
			P2P   & Torrent & 50000 \\
			\midrule
			\multirow{4}[2]{*}{Streaming} & Netflix & \multirow{4}[2]{*}{50000} \\
				  & Spotify &  \\
				  & Vimeo &  \\
				  & Youtube &  \\
			\midrule
			\multirow{4}[2]{*}{VoIP} & Facebook\_audio & \multirow{4}[2]{*}{50000} \\
				  & Hangouts\_audio &  \\
				  & Skype\_audio &  \\
				  & Voipbuster &  \\
			\bottomrule
			\end{tabular}%
	}
	\label{tab:2}%
\end{table}%

In the following, we choose the open UNB ISCX VPN-nonVPN dataset \cite{draper2016characterization} to perform statistical analyses on the byte information of IP packet headers for different categories of network traffic samples. The open UNB ISCX VPN-nonVPN dataset contains both regular traffic and VPN traffic. There are six categories of network traffic data generated by different applications, i.e., Chat, Email, File Transfer, P2P, Streaming, and VoIP (seven in the official description document, but only six in the actual downloaded files). The open UNB ISCX VPN-nonVPN dataset is unbalanced. In order to avoid the impact of an unbalanced dataset on the experimental results, we construct a balanced dataset with 50000 samples for each traffic category. Details of the ISCX dataset are shown in Table 2 and is called ISCX dataset in this paper for simplicity.

For the original Pcap files of each application, we employ the Dpkt Library in Python for data preprocessing. Based on the experimental requirements of this paper, we extract a total of 20 bytes of IP layer header data for each packet and remove the IP address. Table 3 shows IP header values for some samples in the ISCX dataset. Here, the label field represents the category of network traffic, 0-5 respectively represents Chat, Email, File Transfer, P2P, Streaming, and VoIP, and we mark the corresponding traffic category for each sample.

\begin{table*}[htbp,h]
	\normalsize
	\centering
	\caption{Samples of IP Packet header byte values and label for six traffic types on ISCX dataset}
	\scalebox{0.7}{
		\begin{tabular}{cccccccccccccc}
			\toprule
			Traffic Type & Byte1 & Byte2 & Byte3 & Byte4 & Byte5 & Byte6 & Byte7 & Byte8 & Byte9 & Byte10 & Byte11 & Byte12 & Label \\
			\midrule
			Chat  & 69    & 0     & 4     & 143   & 108   & 209   & 64    & 0     & 128   & 6     & 3     & 94    & 0 \\
			Email & 69    & 0     & 5     & 220   & 90    & 160   & 64    & 0     & 32    & 6     & 101   & 46    & 1 \\
			File Transfer & 69    & 0     & 0     & 72    & 75    & 84    & 64    & 0     & 34    & 6     & 46    & 146   & 2 \\
			P2P   & 69    & 0     & 0     & 40    & 100   & 20    & 64    & 0     & 128   & 6     & 25    & 118   & 3 \\
			Streaming & 69    & 0     & 0     & 52    & 129   & 17    & 64    & 0     & 64    & 6     & 145   & 40    & 4 \\
			VoIP  & 69    & 0     & 0     & 211   & 172   & 169   & 64    & 0     & 76    & 6     & 251   & 65    & 5 \\
			\bottomrule
			\end{tabular}%
	}
	\label{tab:3}%
\end{table*}%

As shown in Table 3, we can get all the values of each field and the corresponding traffic category for every sample. The raw data of network traffic can be divided into bytes, which can be transformed into decimal numbers within the range of 0-255. Therefore, we first normalized the 12-byte packet header data by dividing each byte by 255. This ensures that the values of each byte are within the range of 0 to 1. Next, we divided the range of 0 to 1 into 20 intervals, with a distance of 0.05 between each interval. Except for the last interval, which is closed, the other 19 intervals are left-closed and right-open. For each byte, we map the values of 50,000 samples in the dataset to the 20 intervals and subsequently count the number of samples in each interval. This process allows us to obtain the value distribution of the 12-byte packet header information for different types of traffic in the dataset.

Based on the statistical results, we plotted a three-dimensional bar chart, as shown in Fig. 2. Due to the varying lengths of different fields within the IP packet header, being separated by bytes allows better control of numerical variations within a specific range (0-255), facilitating subsequent data processing. For a fair comparison of performance, it is necessary to consistent the input data with previous works. So far as we know, most of previous works of packet-based network traffic classification separated the IP packet by different bytes \cite{wang2017end,yang2021aefeta,lin2022efficient,wang2022sessionvideo,yang2023network,lotfollahi2020deep,zeng2019test,xu2019traffic,ren2021tree,xie2021self}. Therefore, it is reasonable to separate the IP packet header by bytes in this paper. In Fig. 2, the X-axis (Byte) represents the first to twelfth bytes, the Y-axis (Value) represents 20 intervals ranging from 0 to 1 with an interval of 0.05, and the Z-axis (Quantity) represents the number of occurrences of IP packet header values within each of the corresponding intervals. Different colors are used to represent byte information at different positions.

This dataset consists of only IPV4 network traffic and the network layer header length of a single packet is a fixed 20 bytes. Therefore, Byte 1 is not discriminative for any type of network traffic on this dataset, and the data distribution is the same. For Byte 2, which represents the Priority\&Type of Service, all samples have a preset priority of 000 (Routine) in network transmission. As a result, it also cannot help distinguish different types of network traffic. In the same situation, there is Byte 8, and it can be observed that the values of the Fragment Offset on this dataset are all 0.

For the byte information of other parts, the data distribution will change with the category of network traffic. Bytes 3 and 4 will change with the Total Length of the packet. Byte 7 represents whether different network traffic will be split into fragments during transmission. Similarly, Byte 9 represents Time to Live and Byte 10 represents Protocol, which will also differ. The data distributions of Byte 6 and Byte 12 are relatively wide, with values ranging from 0 to 1, and there is not much difference in the six types of network traffic. However, Byte 6 and Byte 5 are together form Identification, while Byte 12 and Byte 11 are together form Header Checksum. The data distributions of Byte 5 and Byte 11 vary in different categories of network traffic, so overall, they still have a certain degree of differentiation. Among these, the Total Length and Protocol can directly reflect the packet's intrinsic characteristics, as packets generated by different applications/protocols typically have different lengths and use different transport protocols. The effect of Total Length on network traffic classification is also demonstrated in some flow-based works \cite{liu2019fs,shapira2021flowpic,babaria2021flowformers,yang2023network}.

In summary, the statistical characteristics of numerical distributions reveal that the IP packet header does possess some discriminative features for different traffic types. However, these features are low-dimensional, so, it's necessary to employ deep learning methods to further explore high-dimensional features within the packet header information that aid in distinguishing different traffic types and achieving more precise classification. Due to space limitations, this section only analyzes the statistical distribution of IP packet header features on the ISCX dataset. In addition, we conduct experimental validation of the effectiveness of the network traffic classification scheme based on the IP packet header using a merged dataset in chapter 5. The experimental results show that this scheme can achieve a classification accuracy of over 95\%. For more details, please refer to Section 5.4.
	
\section{Methodology}

In this paper, we propose a mixed model for online network traffic classification based on external attention and convolution. As shown in Fig. 3, it consists of five modules, i.e., data preprocessing, the embedding layer, the external attention layer, the convolutional layer, and the linear layer. According to the above scheme based on IP packet header, data preprocessing is responsible for converting the raw data from the Pcap files to byte data, which is then put into the embedding layer. The embedding layer serves two purposes. On the one hand, it adds the field position information into its mathematical value to distinguish different fields with the same values. On the other hand, it maps the low-dimensional byte data into the higher one for digging more complex classification information. The high-dimensional data is then put into the deep learning part, including the external attention layer, convolution layer, and linear layer. The purpose of the external attention mechanism is to further enhance the helpful information and weaken interference for network traffic classification, which are contained in the high-dimensional outputs of the embedding layer. In other words, it is to further strengthen the important features of the classification task. The high-dimensional features are then further learned by the convolution layer to capture the significant byte and packet-level semantics. Finally, the linear layer is applied to make the decision for the corresponding network traffic category.

\begin{figure*}[htbp]
	\centerline{\includegraphics[scale=0.6]{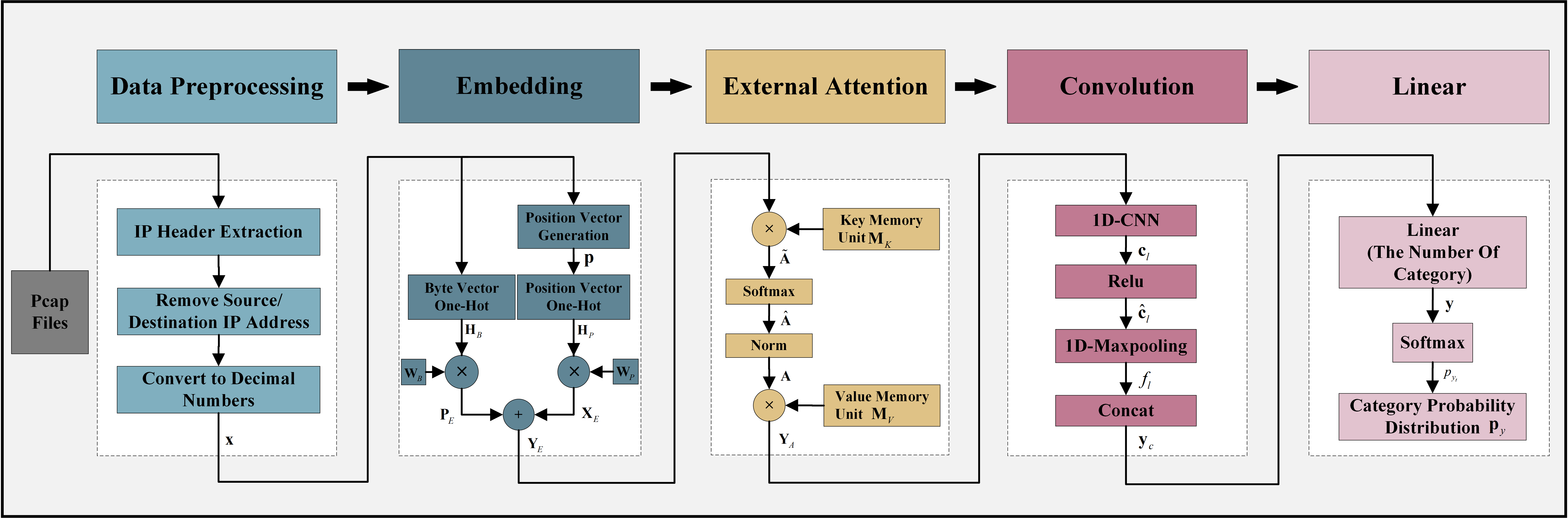}}
	\caption{An external attention and convolution mixed model by IP packet header for network traffic classification, mainly including data preprocessing, embedding layer, external attention layer, convolutional layer, and linear layer.} 
	\label{fig:3} 
\end{figure*}

\subsection{Data Preprocessing}
The IP packets captured from networks are usually saved in hexadecimal format in the Pcap file. First, we extract IP packet header data from the Pcap files. And then, the source and destination IP addresses are eliminated and the 12-byte IP packet header data is achieved by the IP header compression module. Based on the analyses in Section 3, the 12-byte IP packet header is taken as a sample and assigned a label value of its corresponding network traffic category. Finally, the hexadecimal header data is converted to decimal in bytes, and each byte corresponds to a number in [0, 255] through the byte vector generation module. After the above preprocessing, each packet forms a sequence of numbers which is the input vector \(\textbf{x} \in {\mathbb{R}^{^{N \times 1}}}{\text{(}}N{\text{ = 12)}}\) of the deep learning model.

\subsection{Embedding layer}
The bytes in different positions of the IP packet header represent different fields. In other words, the bytes in different fields have different physical meanings and may correspond to different traffic categories, although they have the same value. Therefore, the position information of each byte needs to be integrated on the basis of the input vector \(\textbf{x}\). Here, we use the value 0 to \(N - 1\) to represent the position information from the first byte to the \(N{\rm{th}}\) byte. And hence, the position information can be described by a vector \(\textbf{p}\), where \(\textbf{p}\)=[0,1,2,...,\(N - 1\)]. In this paper, we choose the embedding layer to map the one-dimensional input vector \(\textbf{x}\) and its position information \(\textbf{p}\) to higher dimensions. In order to minimize the computational complexity, this paper selects a classical method, i.e., Word2Vec \cite{mikolov2013efficient,pennington2014glove,joulin2016bag}, to make embedding. For \(N\) bytes of the input vector of \(\textbf{x} = ({x_1}, \cdots ,{x_n}, \cdots ,{x_N})\), each byte is first mapped in a high-dimensional way using One-Hot encoding. Since the decimal number of each byte is in the range [0, 255], the One-Hot encoding has a dictionary size of 256. Thus, the input data \({x_n}\) is mapped into a vector \({\textbf{h}_{B,n}} = (0, \cdots ,1, \cdots ,0)\). Here, \(\textbf{h}_{B,n}\) is a \(1 \times 256\) vector with only the \({x_n}th\) element taking 1 and the rest elements taking 0. Thus, after One-Hot encoding, the input data \(\textbf{x}\) is mapped to a matrix of \(\textbf{H}_{B}\), where \(\textbf{H}_{B} = {(\textbf{h}_{B,1}, \cdots ,\textbf{h}_{B,n}, \cdots ,\textbf{h}_{B,N})^T}\). The matrix \(\textbf{H}_{B}\) is then passed through a fully-connected neural network layer, and the final output embedding matrix is denoted as \({\textbf{X}_E} \in {\mathbb{R}^{N \times D}}\), where \(D\) is a hyperparameter representing the predefined output dimension of the word embedding layer. Thus, the mathematical expression of \({\textbf{X}_E}\) is given by

\begin{equation}\label{eqn-1}
	{\textbf{X}_E} = {\textbf{H}_B} \times {\textbf{W}_B},
\end{equation}where \({\textbf{W}_B} \in {\mathbb{R}^{256 \times D}}\) is the weight matrix of the fully-connected layer.

For the position vector \(\textbf{p} = ({p_1}, \cdots ,{p_n}, \cdots ,{p_N})\), the One-Hot encoding for the \(N{\rm{th}}\) byte is represented as \({\textbf{h}_{P,n}} = (0, \cdots ,1, \cdots ,0)\), where \({\textbf{h}_{P,n}}\) is a \(1 \times 12\) vector with only the \({p_n}th\) element taking 1 and the rest elements taking 0. Then, the position vector \(\textbf{p}\) is mapped to a matrix of \(\textbf{H}_{P}\), where \(\textbf{H}_{P} = {(\textbf{h}_{P,1}, \cdots ,\textbf{h}_{P,n}, \cdots ,\textbf{h}_{P,N})^T}\). The same embedding operation as the input vector \(\textbf{x}\) is used to obtain the position Embedding matrix \(\textbf{P}_E \in {\mathbb{R}^{N \times D}}\) and 
\begin{equation}\label{eqn-2}
	{\textbf{P}_E} = {\textbf{H}_P} \times {\textbf{W}_P},
\end{equation}where \({\textbf{W}_P} \in {\mathbb{R}^{12 \times D}}\) is another weight matrix of the fully-connected layer.

Finally, the high-dimensional byte information and its position information are added to obtain the output matrix of embedding layer. And the output \(\textbf{Y}_E\) of data preprocessing is described as,

\begin{equation}\label{eqn-3}
	\textbf{Y}_E = \textbf{X}_E + \textbf{P}_E,
\end{equation}where \(\textbf{Y}_E \in {\mathbb{R}^{N \times D}}\).

\subsection{External Attention}
In this paper, we adopt a novel external attention-based mechanism in \cite{guo2021beyond} to further enhance the classification features. Instead of each input sample corresponding to a different key matrix and value matrix in the traditional self-attention, all samples in the dataset share the same key matrix and value matrix, which are called the external memory units in external attention. Therefore, the operations of three linear layers in the self-attention mechanism are eliminated, which are used to generate the query matrix, key matrix, and value matrix for each sample. Thus, the temporal computational complexity of external attention can be reduced. In addition, since the two external memory units are the optimalities on a statistical average of the whole dataset, it can learn the most discriminative features and capture the most informative part for all samples, as well as play a certain role in regularization and have better generalization.

In this paper, the two external memory units in external attention are denoted as \({\textbf{M}_K}\) and \({\textbf{M}_V}\), where they are both \(S \times D\) matrixes. \(S\) and \(D\) are the predefined hyperparameters, and \(D\) is consistent with the embedding dimension in the embedding layer. In the first step, we obtain the attention score matrix by the following equations, multiplying the embedding data \({\textbf{Y}_E}\) with the transpose of the matrix \({\textbf{M}_K}\) to obtain the matrix \(\tilde {\textbf{A}} \in {\mathbb{R}^{N \times S}}\),

\begin{equation}\label{eqn-4}
	\tilde {\textbf{A}} = {({{\tilde a}_{i,j}})_{N \times S}} = {\textbf{Y}_E} \times {\textbf{M}}_K^T.
\end{equation}Then, the softmax operation is done by the following function,

\begin{equation}\label{eqn-5}
	\hat {\textbf{A}} = {({{\hat a}_{i,j}})_{N \times S}} = {({\rm{softmax(}}{{\tilde a}_{i,j}}{\rm{))}}_{N \times S}}.
\end{equation}Next, normalization is done to obtain the attention score matrix as follows,

\begin{equation}\label{eqn-6}
	{\textbf{A}} = {({a_{i,j}})_{N \times S}} = {\left( {{{{{\hat a}_{i,j}}} \mathord{\left/
 {\vphantom {{{{\hat a}_{i,j}}} {\sum\limits_{j = 1}^S {{{\hat a}_{i,j}}} }}} \right.
 \kern-\nulldelimiterspace} {\sum\limits_{j = 1}^S {{{\hat a}_{i,j}}} }}} \right)_{N \times S}}.
\end{equation}Finally, the attention score matrix \(\textbf{A}\) is multiplied with the matrix \({\textbf{M}_V}\) to obtain the output matrix \({\textbf{Y}_A} \in {\mathbb{R}^{N \times D}}\) of the external attention layer according to Eq. (7),

\begin{equation}\label{eqn-7}
		\begin{array}{l}
			{\textbf{Y}_A} = \left( \begin{array}{l}
			y_{1,1}^{(A)}\quad y_{1,2}^{(A)}\quad  \cdots \quad y_{1,D}^{(A)}\\
			y_{2,1}^{(A)}\quad y_{2,2}^{(A)}\quad  \cdots \quad y_{2,D}^{(A)}\\
			\;\,\, \vdots \quad \quad \;\, \vdots \quad \;\; \ddots \quad \;\;\; \vdots \\
			y_{N,1}^{(A)}\quad y_{N,2}^{(A)}\;\;\; \cdots \quad y_{N,D}^{(A)}
			\end{array} \right)\\
			 = {({\textbf{y}}_1^{(A)}, \cdots ,{\textbf{y}}_n^{(A)}, \cdots ,{\textbf{y}}_N^{(A)})^T} = {\textbf{A}} \times {\textbf{M}_V}.
			\end{array}
\end{equation}Here, it is worth noting that the process of achieving the attention scores and output matrix \({\textbf{Y}_A}\) mainly based on each byte data. Therefore, the external attention mechanism can be seen as the enhancement of the high-dimensional byte-level features to describe more important information in the task of classifying network service traffic.

\subsection{Convolutional Layer}
\begin{figure}[htbp]
	\centerline{\includegraphics[scale=0.4]{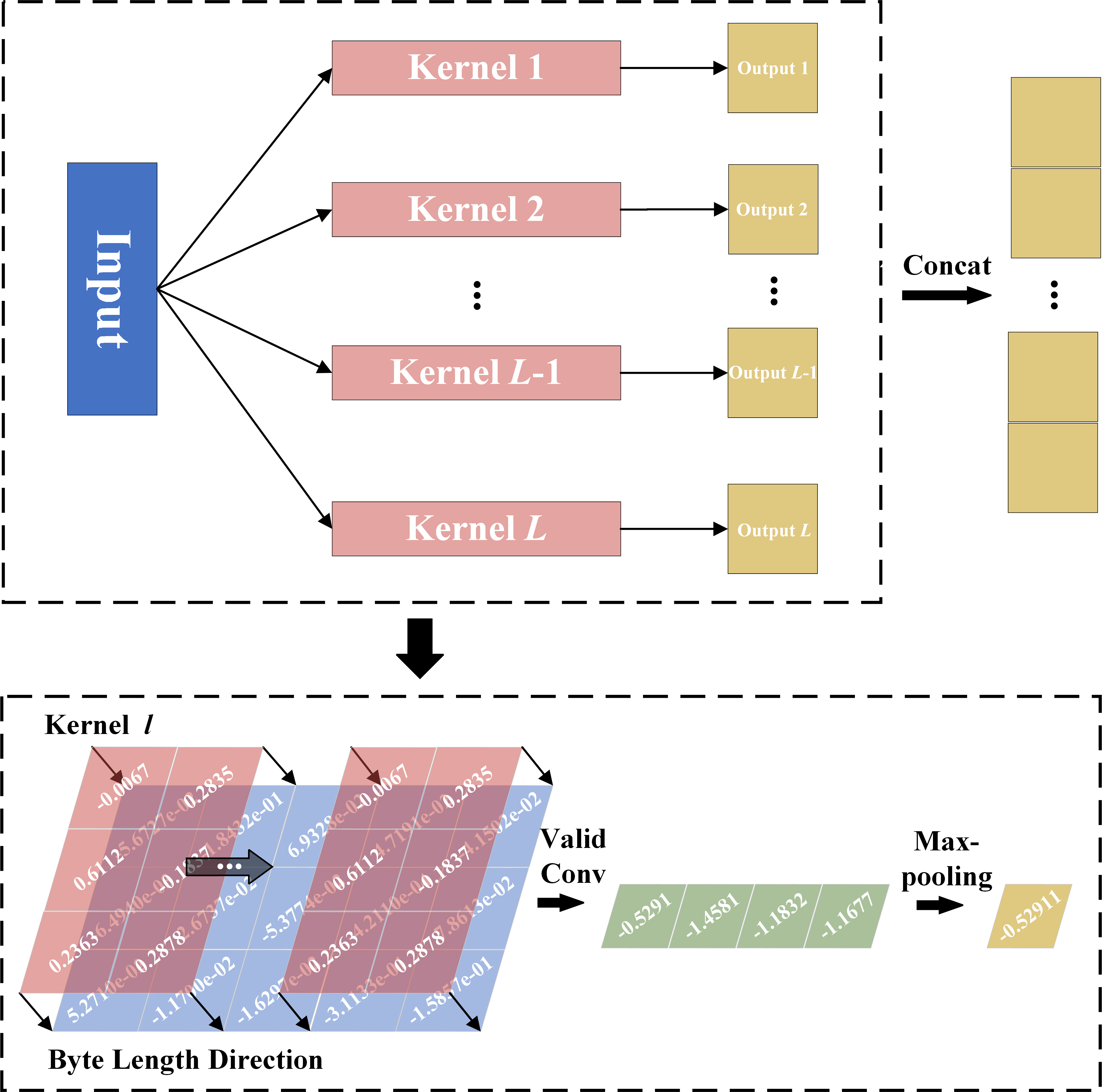}}
	\caption{One-dimensional valid convolution and concatenation operation} 
	\label{fig:4} 
\end{figure}

After the external attention layer, we use the convolutional layer to capture the high-dimensional features of the packet headers that contain significant byte and packet-level semantics. Just as the achievements in section 4.2, both the value of each byte and their relationships are meaningful for network traffic classification tasks, which is similar to the phenomenon that both each word’s meaning and their order are important for the sentence semantics in the natural language domain. Therefore, as shown in Fig. 4, the convolution should be made in the byte length direction to capture their relationships. 

In addition, the sparse interaction and parameter-sharing properties of convolutional operations can be utilized to compress the dimensionality of the data and finally save time for later operations of the model, which is also important for online tasks. Compared with high-dimensional convolution, one-dimensional convolution can also reduce computational complexity and is more suitable for online service traffic classification scenarios. Considering the above two factors, we adopt the most basic valid one-dimensional convolution to convolve the input matrix \({\textbf{Y}_A}\) in the byte direction.

Since the features extracted by different convolution kernels are different, this paper uses multiple convolution kernels \({\textbf{K}_l} \in {\mathbb{R}^{Q \times D}}(l = 1,2, \cdots ,L)\) at the same time to achieve feature maps. On each convolution kernel, we perform a one-dimensional valid convolution operation in the dimension of bytes, and the specific operation of convolution is shown in Figure 4. Thus, the convolution result \({\textbf{c}_l} \in {\mathbb{R}^{(N - Q + 1) \times 1}}\) can be derived from Eqs. (8)-(10),

\begin{equation}\label{eqn-8}
	\begin{gathered}
		{{\tilde {\textbf{C}}}_{l,u}} = \left[ \begin{gathered}
		\tilde c_{1,1}^{(l,u)}\quad \tilde c_{1,2}^{(l,u)}\quad  \cdots \quad \tilde c_{1,Q}^{(l,u)} \hfill \\
		\tilde c_{2,1}^{(l,u)}\quad \tilde c_{2,2}^{(l,u)}\quad  \cdots \quad \tilde c_{2,Q}^{(l,u)} \hfill \\
		\;\;\, \vdots \quad \quad \;\; \vdots \quad \;\;\; \ddots \quad \;\;\, \vdots  \hfill \\
		\tilde c_{D,1}^{(l,u)}\quad \tilde c_{D,2}^{(l,u)}\quad  \cdots \quad \tilde c_{D,Q}^{(l,u)} \hfill \\ 
	  \end{gathered}  \right] \hfill \\
		 = {\textbf{Y}}_A^T(:,u:u + Q - 1) * {\textbf{K}_l},u \in [1,N - Q + 1] \hfill, \\
	  \end{gathered} 
\end{equation}

\begin{equation}\label{eqn-9}
	{c_{l,u}} = \sum\limits_{i = 1}^D {\sum\limits_{j = 1}^Q {\tilde c_{i,j}^{(l,u)}} },
\end{equation}

\begin{equation}\label{eqn-10}
	{\textbf{c}_l} = {[{c_{l,1}}, \cdots ,{c_{l,u}}, \cdots ,{c_{l,N - Q + 1}}]^T}.
\end{equation}

In order to increase the nonlinear expression ability of the convolution layer, the Relu function is chosen in this paper to make nonlinear changes to the above convolution result \({\textbf{c}_l}\) and obtain the transformed feature map \({{\hat {\textbf{c}}}_l}\), which is mathematically described as follows,

\begin{equation}\label{eqn-11}
	{{\hat {\textbf{c}}}_l} = {[{{\hat c}_{l,1}}, \cdots ,{{\hat c}_{l,u}}, \cdots ,{{\hat c}_{l,N - Q + 1}}]^T} = {\text{Relu}}({\textbf{c}_l}).
\end{equation}Then, the max-pooling is employed to extract the maximum value from the feature map \({{\hat {\textbf{c}}}_l}\). The feature map extracted from the \(l{\rm{th}}\) channel is down-sampled to obtain the unique value \({f_l}\) with the most important feature. The mathematical formula is as follows,

\begin{equation}\label{eqn-12}
	\begin{array}{l}
		{f_l} = \max ({{\hat {\textbf{c}}}_l})\\
		 = \max ({{\hat c}_{l,1}}, \cdots ,{{\hat c}_{l,u}}, \cdots ,{{\hat c}_{l,N - Q + 1}}),l = 1,2, \cdots L.
		\end{array}
\end{equation}Finally, the results on each convolution kernel are concatenated together to form a one-dimensional convolutional eigenvector \({\textbf{y}_c} \in {\mathbb{R}^{L \times 1}}\), where \({\textbf{y}_c} = {[{f_1}, \cdots ,{f_l}, \cdots ,{f_L}]^T}\).

\subsection{Linear Layer and Output}
After the data passes through the convolution layer, we need to map the convoluted low-dimensional features to the specific network traffic category as the final output of the model. Assuming that the total number of traffic categories is \(T\). Firstly, a full connection layer is applied to match the dimension of \({\textbf{y}_c}\) to \(T\) and get a \(T \times 1\) vector \(\textbf{y}\) according to the following equation,

\begin{equation}\label{eqn-13}
	\textbf{y} = [{y_1}, \cdots ,{y_t}, \cdots ,{y_T}] = {\textbf{W}}' \times {{\textbf{y}}_c}.
\end{equation}Where \({\textbf{W}}' \in {\mathbb{R}^{T \times L}}\) is the weight matrix of the linear layer. Then, the probability distribution of the current input sample belonging to category \(t\) is calculated by the softmax activation function as follows,

\begin{equation}\label{eqn-14}
	{p_{{y_t}}} = {\rm{softmax}}({y_t}) = \frac{{{e^{{y_t}}}}}{{\sum\limits_{t = 1}^T {{e^{{y_t}}}} }},t = 1,2, \ldots T.
\end{equation}Therefore, the probability distribution that the current input sample belongs to each category can be expressed as \({\textbf{p}_y} = {[{p_{{y_1}}}, \cdots ,{p_{{y_t}}}, \cdots {p_{{y_T}}}]^T}\).

After obtaining the final predicted category, we compare it with the true category probability of the current sample \({\textbf{p}_x} = {[{p_{{x_1}}}, \cdots ,{p_{{x_t}}}, \cdots {p_{{x_T}}}]^T}\) to calculate the most common cross-entropy loss function by Eq. (15), 

\begin{equation}\label{eqn-15}
	{\rm{Loss}} =  - \sum\limits_{t = 1}^T {{p_{{y_t}}}\log } {p_{{x_t}}}.
\end{equation}

Finally, the model is back-propagated based on the current overall loss, and the parameters are updated and optimized by the gradient descent method until the loss is minimized or a set threshold is reached.

\section{Experiments}
\subsection{Experimental Environment and Evaluation}
\emph{\textbf{Experimental environment:}} We choose the deep learning framework Pytorch to conduct our experiments. The server GPU is NVIDIA GeForce RTX 2080Ti, the Python version is 3.8.8, and Python libraries such as Dpkt, Numpy, Pandas, and Scikit-learn are also used.

\emph{\textbf{Performance metrics:}} The classical performance metrics for measuring classification results are mainly accuracy, precision, recall, and F1-score \cite{xie2021self}. Accuracy is defined as the ratio of the number of samples correctly classified by the model to the total number of samples and is used to measure the judgment ability of the model on all samples. Precision refers to the percentage of real positive samples among the total predicted positive samples by the model. Recall is the percentage of the predicted positive samples in the total correctly predicted samples. The F1-score is used to comprehensively measure the recall rate and accuracy of the model. These four metrics are defined as shown in Eqs. (16)-(19),

\begin{equation}\label{eqn-16}
	{\rm{Accuracy}} = \frac{{{\rm{TP + TN}}}}{{{\rm{TP + TN + FP + FN}}}},
\end{equation}

\begin{equation}\label{eqn-17}
	{\rm{Precision}} = \frac{{{\rm{TP}}}}{{{\rm{TP + FP}}}},
\end{equation}

\begin{equation}\label{eqn-18}
	{\rm{Recall}} = \frac{{{\rm{TP}}}}{{{\rm{TP + FN}}}},
\end{equation}

\begin{equation}\label{eqn-19}
	{\rm{F1 - score}} = \frac{{{\rm{2 \times Recall \times Precision}}}}{{{\rm{Recall + Precision}}}}.
\end{equation}Where TP indicates the number of samples that the true value is a positive class and the predicted value is also a positive class; FN indicates the number of samples that the true value is a positive class and the predicted value is a negative class; FP indicates the number of samples that the true value is a negative class while the predicted value is a positive class; TN indicates the number of samples that the true value is a negative class while the predicted value is a negative class. Generalized to multiple classifications, the positive class represents the category to which the sample truly belongs, while the negative classes represent all other categories.

\begin{figure*}[htbp]
	\centerline{\includegraphics[scale=0.6]{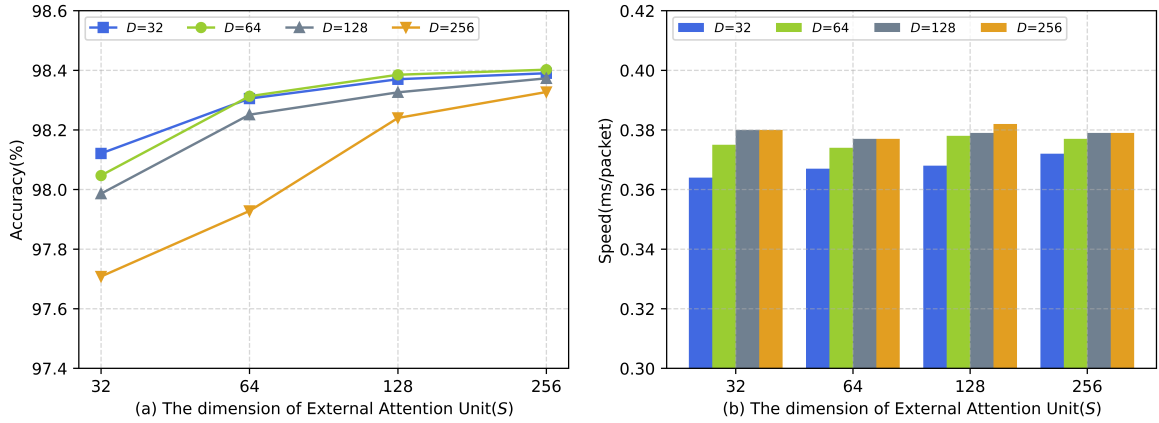}}
	\caption{Performance comparisons of hypeparameter \(S\) and \(D\) with different values on ISCX dataset} 
	\label{fig:5} 
\end{figure*}

\subsection{Hyperparameter Selections}

In this paper, we set the optimizer to Adam, the learning rate to 0.001, the probability of Dropout to 0.1, the convolutional kernel size to 3, the Batch Size to 128, and the epochs of training and test to 200. Note that the input data lengths for all experiments of the method are 12 bytes. As described in Section 4, \(S\) and \(D\) are the hyperparameters of the embedding and the external attention layers. In this paper, we evaluate and determine the values of \(S\) and \(D\) by means of experiments.

Regarding the classification accuracy, as shown in Figure 5(a), it can be observed that with the increase of parameter \(S\), the accuracy demonstrates an overall upward trend. However, when \(S\) is increased from 128 to 256, the accuracy improvement stabilizes, gradually converging around 98.4\%, and the increment in accuracy fluctuates within a small range, indicating no significant improvement. As for the dimension of the Embedding layer \(D\), it can be observed that irrespective of the value of \(S\), the classification accuracy is higher when \(D\) is set to 32 or 64 compared to when it is set to 128 or 256. This suggests that, in this experiment, a larger embedding dimension is not necessarily better. Continuous increment of \(D\) actually leads to a decrease in accuracy.

In terms of classification speed (i.e., model inference time), due to hardware limitations, the inference speed may exhibit minor fluctuations and not remain constant. Therefore, we consider the average speed under different hyperparameters. As illustrated in Figure 5(b), Experimental results reveal that parameter \(S\) is primarily associated with computations related to the external attention layer, thus, its variations have a relatively minor impact on the final classification speed. Conversely, the dimension of \(D\), affects computations at each layer of the model, leading to slight variations in classification speed for different \(D\) values. Furthermore, although increasing the values of hyperparameters \(S\) and \(D\) may increase the number of model parameters, their relationship with the model computation is not strictly linear. The reason may be that there is no direct causal relationship between the amount of model computation and the speed of model inference according to the roofline model in \cite{ding2019instruction}. Actual measurement is the most accurate way to evaluate performance. 

In summary, considering both the overall classification accuracy and the classification speed, we choose the values of \(S\) and \(D\) corresponding to higher accuracy and shorter time consumption. It is reasonable to select \(S\) with 128 and \(D\) with 32. Therefore, in the following experiments, the hyperparameters \(S\) and \(D\) are set as the above conclusion.

\subsection{Baselines Comparisons}

\emph{\textbf{Baselines:}} The proposed external attention and convolution mixed (ECM) model is compared with four other packet-based deep learning methods, i.e., CNN \cite{lotfollahi2020deep}, LSTM \cite{xu2019traffic}, CNN+LSTM and SAM \cite{xie2021self}. The reasons for choosing them are as follows. As the detailed descriptions in section 2, CNN and LSTM are classical methods for network traffic classification, and CNN+LSTM is the fusion of these two models. SAM adopts the self-attention mechanism for network traffic classification tasks, which is the most similar to ours. 

\begin{table}[htbp]
	\normalsize
	\centering
	\caption{Experimental settings}
	\scalebox{0.6}{
	  \begin{tabular}{cc}
	  \toprule
	  Methods & Settings \\
	  \midrule
	  CNN   & kernel\_size=3, kernels=200 \\
	  \midrule
	  LSTM  & hidden\_layer\_size=[64,64,32] \\
	  \midrule
	  CNN+LSTM & the same as CNN \& LSTM \\
	  \midrule
	  SAM   & embedding\_dim=256, kernel\_size=[3,4], kernels=256 \\
	  \midrule
	  ECM   & \(S\)=128, \(D\)= 32, kernel\_size=3, kernels=256 \\
	  \midrule
	  \multicolumn{2}{c}{Other settings: trainging batch\_size = 128, epoch is 200, optimizer is Adam, learning rate is 0.001.} \\
	  \bottomrule
	  \end{tabular}%
	}
	\label{tab:4}%
  \end{table}%

In this subsection, we make some performance comparisons of the proposed ECM model with some baselines. As for the settings of hyperparameters for the specific model structure (such as the size of convolution kernels and the dimension of hidden layers), we keep consistent with the original works. The experimental settings for each method in the baselines are shown in Table 4.

\begin{table}[htbp]
	\normalsize
	\centering
	\caption{Performance comparisons of baselines on ISCX dataset}
	\scalebox{0.7}{
		\begin{tabular}{cccccc}
			\toprule
				  & Acc.  & Pre.  & Rec.  & F1.  & Speed(ms/packet) \\
			\midrule
			CNN   & 97.75 & 97.76 & 97.75 & 97.76 & 0.38 \\
			\midrule
			LSTM   & 96.42 & 96.45 & 96.42 & 96.43 & 0.35 \\
			\midrule
			CNN+LSTM   & 97.60 & 97.60 & 97.60 & 97.60 & 0.69 \\
			\midrule
			SAM   & 97.07 & 97.09 & 97.07 & 97.08 & 0.68 \\
			\midrule
			\textbf{ECM}   & \textbf{98.39} & \textbf{98.39} & \textbf{98.39} & \textbf{98.39} & \textbf{0.36} \\
			\bottomrule
			\end{tabular}%
	}
	\label{tab:5}%
\end{table}%

As shown in Table 5, based on the average accuracy from ten-fold cross validation, ECM achieves the highest accuracy, precision, recall, and F1-score (all exceeding 98\%). Moreover, ECM achieves an inference time of milliseconds per individual data packet, making it suitable for real-time classification needs. Even though the LSTM model has the fastest classification speed, our ECM is only approximately 0.01 milliseconds slower than LSTM for the classification of individual data packets, while the classification speed of other methods is slower than that of ECM. The difference is within the speed fluctuation range (0.05ms-0.1ms). And in the other dataset as described in the following subsection 5.4, the speeds of both these two methods are both 0.35ms. Furthermore, compared with the previous works for online network traffic classification \cite{ren2021tree,xie2020sam,xie2021self}, ECM achieves the same level of performance, i.e., millisecond-level classification speed. Therefore, in a comprehensive manner, our proposed ECM model based on the IP packet header has an excellent performance in terms of accuracy and speed.

\subsection{Verification of IP Packet Header Method on Another Dataset}

\begin{table}[htbp,h]
	\normalsize
	\centering
	\caption{BUPD Dataset categories and numbers of samples}
	\scalebox{0.7}{
		\begin{tabular}{ccc}
			\toprule
			Traffic Type & Content & Quantity \\
			\midrule
			\multirow{2}[2]{*}{Chat} & QQ    & \multirow{2}[2]{*}{50000} \\
				  & Wechat &  \\
			\midrule
			\multirow{2}[2]{*}{Email} & Outlook & \multirow{2}[2]{*}{50000} \\
				  & Mymail &  \\
			\midrule
			File Transfer & SMB   & 50000 \\
			\midrule
			\multirow{3}[2]{*}{P2P} & Bitcomet & \multirow{3}[2]{*}{50000} \\
				  & Thunder &  \\
				  & Bittorrent &  \\
			\midrule
			\multirow{12}[2]{*}{Streaming} & IQIYI & \multirow{12}[2]{*}{50000} \\
				  & NeteaseClouodMusic &  \\
				  & QQmusic &  \\
				  & TencentVideo &  \\
				  & Youku &  \\
				  & Bilibili &  \\
				  & HuyaLive &  \\
				  & KugouMusic &  \\
				  & MangoTV &  \\
				  & PPTV  &  \\
				  & SouhuVideo &  \\
				  & TikTok &  \\
			\midrule
			\multirow{9}[2]{*}{Game} & CrossFire & \multirow{9}[2]{*}{50000} \\
				  & LeagueOfLegends &  \\
				  & MineCraft &  \\
				  & WorldOfWarcraft &  \\
				  & DOTA  &  \\
				  & Hepingjingying &  \\
				  & Huangyexingdong &  \\
				  & Huoyingrenzhe &  \\
				  & Juediqiusheng &  \\
			\midrule
			\multirow{2}[2]{*}{Meeting} & Voovmeeting & \multirow{2}[2]{*}{50000} \\
				  & TencentMeeeting &  \\
			\midrule
			\multirow{7}[2]{*}{Web} & AmazonWithChrome & \multirow{7}[2]{*}{50000} \\
				  & DoubanWithChrome &  \\
				  & TaobaoWithChrome &  \\
				  & SinaWeibo &  \\
				  & Hongxiutianxiang &  \\
				  & QidianCNWeb &  \\
				  & Xiaoxiangshuyan &  \\
			\bottomrule
			\end{tabular}%
	}
	\label{tab:6}%
\end{table}%

\begin{table}[htbp]
	\normalsize
	\centering
	\caption{Performance comparisons of baselines on BUPD dataset}
	\scalebox{0.7}{
		\begin{tabular}{cccccc}
			\toprule
				  & Acc.  & Pre.  & Rec.  & F1.  & Speed \\
			\midrule
			CNN   & 92.54 & 92.98 & 92.55 & 92.76 & 0.38 \\
			\midrule
			LSTM   & 90.74 & 92.44 & 90.75 & 91.59 & 0.35 \\
			\midrule
			CNN+LSTM   & 92.91 & 93.25 & 92.92 & 93.08 & 0.69 \\
			\midrule
			SAM   & 92.76 & 92.94 & 92.76 & 92.85 & 0.68 \\
			\midrule
			\textbf{ECM}   & \textbf{95.57} & \textbf{95.62} & \textbf{95.57} & \textbf{95.59} & \textbf{0.35} \\
			\bottomrule
			\end{tabular}%
	}
	\label{tab:7}%
\end{table}%

To further validate the efficacy of our proposed ECM model, which exclusively utilizes the network layer IP packet header for network traffic classification, we conduct experiments in this section using an additional dataset. Throughout these experiments, we maintain consistent experimental settings in section 5.1 and 5.3, as well as the hyperparameter settings in section 5.2. This merged dataset is a combination of multiple datasets, including open-source dataset (Beijing University of Aeronautics and Astronautics, BUAA) \cite{wang2022sessionvideo} and (University of Science and Technology of China, USTC) \cite{wang2017malware}, as well as some \textbf{P}rivate \textbf{D}ata collected by ourselves (we call this merged dataset BUPD in this paper). It contains 38 different applications/protocols, as detailed in Table 6. We have categorized them into 8 major network traffic types: Chat, Email, File Transfer, P2P, Streaming, Game, Meeting, and Web. Similar to the ISCX dataset, we select 50,000 samples for each traffic type to maintain dataset balance.

As shown in Table 5 and 7, it can be observed that the performances on BUPD data is worse than that in ISCX. The reason may be the number of traffic types and the applications/protocols on BUPD is more than that on ISCX, which are respectively shown in Table 2 and Table 6. That is to say as the number of traffic types and the applications/protocols increase, the overall classification accuracy tends to decrease. However, our proposed ECM still achieves the highest accuracy, precision, recall, and F1-score (all exceeding 95\%) on BUPD dataset. Furthermore, with the increasing diversity of traffic and application categories, the gap between ECM and other models continues to widen, which means that the classification ability of our proposed ECM is better than baselines. The results on both two datasets demonstrate the utility of using only the network layer IP header information for distinguishing network traffic and highlight the strong performance of our proposed ECM in terms of both classification accuracy and classification speed.

\begin{table}[htbp]
	\normalsize
	\centering
	\caption{Performance comparisons of ECM with different input lengths on ISCX dataset}
	\scalebox{0.7}{
		\begin{tabular}{cccc}
			\toprule
				  & 12Bytes & 50Bytes & 784Bytes \\
			\midrule
			Accuracy(\%) & 98.39 & 99.58 & 99.61 \\
			\midrule
			Speed(ms/packet) & 0.36 & 0.39 & 0.62 \\
			\bottomrule
			\end{tabular}%
	}
	\label{tab:8}%
\end{table}%

\subsection{Performance Comparisons of ECM with Different Input Lengths}

In this section, to further analyze the performance of our ECM model using 12 bytes as input, we select longer byte lengths for comparison, including 50 bytes as used in paper \cite{xie2021self} and 784 bytes as used in paper \cite{xu2019traffic,ren2021tree}. Apart from the input length, all other experimental settings remain unchanged. The specific accuracy and classification speed results are shown in Table 8. It can be observed that using 50 bytes/784 bytes as input on the ISCX dataset results in a limited improvement in accuracy compared to using only 12 bytes as input. However, even with 50 bytes as input, the accuracy already exceeds 99\%, and further increasing it to 784 bytes did not show significant improvement, as the model reaches its upper limit. However, whether using 50 bytes or 784 bytes, both involve application layer payload and violate user's privacy. As for inference time, increasing the input length leads to an increase in the model's inference speed. When the input length increases from 12 bytes to 50 bytes, the inference time increases by 0.03 milliseconds. However, when using 784 bytes as input, the inference time increases significantly. Therefore, it can be concluded that for traffic-type-level classification granularity, using a smaller amount of packet header information can achieve good classification accuracy, effectively reducing classification latency and avoiding involvement with the application layer payload.

\section{Conclusion and Outlook}
In this paper, we propose a novel external attention and convolution mixed (ECM) model for online network traffic classification tasks. Different from the previous works, we only use the IP packet header to extract the features of network traffic, and no application layer payload information is involved, which can effectively protect privacy. Furthermore, it achieves high classification accuracy using only the 12-byte packet header information, which is further verified by experiments in this paper. The external attention mechanism retains the advantages of interpreting the byte-level semantics of packet headers in the self-attention mechanism but simplifies the internal structure of the self-attention mechanism. And, CNN is also introduced to further capture the packet-level semantics. In other words, the combined of external attention and CNN can exploit both intra-byte and inter-byte information, which is comprehensive for network traffic classification. Therefore, ECM reduces the computational complexity and further improves the classification speed while satisfying the classification accuracy. Experimental analyses show that ECM model with 12-byte IP packet header information has an average speed of about 0.36ms for a single packet with an overall classification accuracy higher than 94.57\%. Compared with the latest works based on the self-attention mechanism, it improves in both accuracy and speed.

However, due to the limitation of the current network public dataset, the network services covered are limited and do not include some new services, such as AR/VR, Internet of Vehicles, etc. Therefore, the future deep learning network traffic classification method based on IP packet headers can target newer service types and use more advanced deep learning model architectures to improve accuracy. For the current six categories of classification tasks, ECM only uses the first 12 bytes to show good performance, but for finer-grained classification tasks, such as application-level classification, whether using only part of the IP header information can also achieve better classification results requires further study.

\bibliographystyle{unsrt}
\bibliography{references}

\begin{thebibliography}{10}

\bibitem{gu2021research}
KG~Yue Gu and D~Li.
\newblock Research on network traffic classification based on machine learning
  and deep learning.
\newblock {\em Telecommun. Sci.}, 37(3):105--113, 2021.

\bibitem{rezaei2019deep}
Shahbaz Rezaei and Xin Liu.
\newblock Deep learning for encrypted traffic classification: An overview.
\newblock {\em IEEE communications magazine}, 57(5):76--81, 2019.

\bibitem{wang2019survey}
Pan Wang, Xuejiao Chen, Feng Ye, and Zhixin Sun.
\newblock A survey of techniques for mobile service encrypted traffic
  classification using deep learning.
\newblock {\em IEEE Access}, 7:54024--54033, 2019.

\bibitem{lotfollahi2020deep}
Mohammad Lotfollahi, Mahdi Jafari~Siavoshani, Ramin Shirali Hossein~Zade, and
  Mohammdsadegh Saberian.
\newblock Deep packet: A novel approach for encrypted traffic classification
  using deep learning.
\newblock {\em Soft Computing}, 24(3):1999--2012, 2020.

\bibitem{zeng2019test}
Yi~Zeng, Zihao Qi, Wencheng Chen, and Yanzhe Huang.
\newblock Test: an end-to-end network traffic classification system with
  spatio-temporal features extraction.
\newblock In {\em 2019 IEEE International Conference on Smart Cloud
  (SmartCloud)}, pages 131--136. IEEE, 2019.

\bibitem{xu2019traffic}
Luyang Xu, Xu~Zhou, Yongmao Ren, and Yifang Qin.
\newblock A traffic classification method based on packet transport layer
  payload by ensemble learning.
\newblock In {\em 2019 IEEE Symposium on Computers and Communications (ISCC)},
  pages 1--6. IEEE, 2019.

\bibitem{ren2021tree}
Xinming Ren, Huaxi Gu, and Wenting Wei.
\newblock Tree-rnn: Tree structural recurrent neural network for network
  traffic classification.
\newblock {\em Expert Systems with Applications}, 167:114363, 2021.

\bibitem{xie2021self}
Guorui Xie, Qing Li, and Yong Jiang.
\newblock Self-attentive deep learning method for online traffic classification
  and its interpretability.
\newblock {\em Computer Networks}, 196:108267, 2021.

\bibitem{wang2017end}
Wei Wang, Ming Zhu, Jinlin Wang, Xuewen Zeng, and Zhongzhen Yang.
\newblock End-to-end encrypted traffic classification with one-dimensional
  convolution neural networks.
\newblock In {\em 2017 IEEE international conference on intelligence and
  security informatics (ISI)}, pages 43--48. IEEE, 2017.

\bibitem{liu2019fs}
Chang Liu, Longtao He, Gang Xiong, Zigang Cao, and Zhen Li.
\newblock Fs-net: A flow sequence network for encrypted traffic classification.
\newblock In {\em IEEE INFOCOM 2019-IEEE Conference On Computer
  Communications}, pages 1171--1179. IEEE, 2019.

\bibitem{shapira2021flowpic}
Tal Shapira and Yuval Shavitt.
\newblock Flowpic: A generic representation for encrypted traffic
  classification and applications identification.
\newblock {\em IEEE Transactions on Network and Service Management},
  18(2):1218--1232, 2021.

\bibitem{zhao2021flow}
Ruijie Zhao, Yiteng Huang, Xianwen Deng, Zhi Xue, Jiabin Li, Zijing Huang, and
  Yijun Wang.
\newblock Flow transformer: A novel anonymity network traffic classifier with
  attention mechanism.
\newblock In {\em 2021 17th International Conference on Mobility, Sensing and
  Networking (MSN)}, pages 223--230. IEEE, 2021.

\bibitem{babaria2021flowformers}
Rushi Babaria, Sharat~Chandra Madanapalli, Himal Kumar, and Vijay Sivaraman.
\newblock Flowformers: Transformer-based models for real-time network flow
  classification.
\newblock In {\em 2021 17th International Conference on Mobility, Sensing and
  Networking (MSN)}, pages 231--238. IEEE, 2021.

\bibitem{yang2021aefeta}
Jingru Yang and Yuanbo Guo.
\newblock Aefeta: Encrypted traffic classification framework based on
  self-learning of feature.
\newblock In {\em 2021 6th International Conference on Intelligent Computing
  and Signal Processing (ICSP)}, pages 876--880. IEEE, 2021.

\bibitem{wu2022online}
Zheng Wu, Yu-ning Dong, Xiaohui Qiu, and Jiong Jin.
\newblock Online multimedia traffic classification from the qos perspective
  using deep learning.
\newblock {\em Computer Networks}, 204:108716, 2022.

\bibitem{li2018byte}
Rui Li, Xi~Xiao, Shiguang Ni, Haitao Zheng, and Shutao Xia.
\newblock Byte segment neural network for network traffic classification.
\newblock In {\em 2018 IEEE/ACM 26th International Symposium on Quality of
  Service (IWQoS)}, pages 1--10. IEEE, 2018.

\bibitem{xie2020sam}
Guorui Xie, Qing Li, Yong Jiang, Tao Dai, Gengbiao Shen, Rui Li, Richard
  Sinnott, and Shutao Xia.
\newblock Sam: self-attention based deep learning method for online traffic
  classification.
\newblock In {\em Proceedings of the Workshop on Network Meets AI \& ML}, pages
  14--20, 2020.

\bibitem{guo2021beyond}
Meng-Hao Guo, Zheng-Ning Liu, Tai-Jiang Mu, and Shi-Min Hu.
\newblock Beyond self-attention: External attention using two linear layers for
  visual tasks.
\newblock {\em arXiv preprint arXiv:2105.02358}, 2021.

\bibitem{lin2022efficient}
Cheng~Yuan Lin, BaiHua Chen, and WeiYao Lan.
\newblock An efficient approach for encrypted traffic classification using cnn
  and bidirectional gru.
\newblock In {\em 2022 2nd International Conference on Consumer Electronics and
  Computer Engineering (ICCECE)}, pages 368--373. IEEE, 2022.

\bibitem{wang2022sessionvideo}
Haiyang Wang, Tongge Xu, Jian Yang, Lijin Wu, and Liqun Yang.
\newblock Sessionvideo: A novel approach for encrypted traffic classification
  via 3d-cnn model.
\newblock In {\em 2022 23rd Asia-Pacific Network Operations and Management
  Symposium (APNOMS)}, pages 1--6. IEEE, 2022.

\bibitem{yang2023network}
Yang Yang, Yu~Yan, Zhipeng Gao, Lanlan Rui, Rui Lyu, Bowen Gao, and Peng Yu.
\newblock A network traffic classification method based on dual-mode feature
  extraction and hybrid neural networks.
\newblock {\em IEEE Transactions on Network and Service Management}, 2023.

\bibitem{aceto2020toward}
Giuseppe Aceto, Domenico Ciuonzo, Antonio Montieri, and Antonio Pescap{\'e}.
\newblock Toward effective mobile encrypted traffic classification through deep
  learning.
\newblock {\em Neurocomputing}, 409:306--315, 2020.

\bibitem{aceto2021distiller}
Giuseppe Aceto, Domenico Ciuonzo, Antonio Montieri, and Antonio Pescap{\'e}.
\newblock Distiller: Encrypted traffic classification via multimodal multitask
  deep learning.
\newblock {\em Journal of Network and Computer Applications}, 183:102985, 2021.

\bibitem{draper2016characterization}
Gerard Draper-Gil, Arash~Habibi Lashkari, Mohammad Saiful~Islam Mamun, and
  Ali~A Ghorbani.
\newblock Characterization of encrypted and vpn traffic using time-related.
\newblock In {\em Proceedings of the 2nd international conference on
  information systems security and privacy (ICISSP)}, pages 407--414, 2016.

\bibitem{mikolov2013efficient}
Tomas Mikolov, Kai Chen, Greg Corrado, and Jeffrey Dean.
\newblock Efficient estimation of word representations in vector space.
\newblock {\em arXiv preprint arXiv:1301.3781}, 2013.

\bibitem{pennington2014glove}
Jeffrey Pennington, Richard Socher, and Christopher~D Manning.
\newblock Glove: Global vectors for word representation.
\newblock In {\em Proceedings of the 2014 conference on empirical methods in
  natural language processing (EMNLP)}, pages 1532--1543, 2014.

\bibitem{joulin2016bag}
Armand Joulin, Edouard Grave, Piotr Bojanowski, and Tomas Mikolov.
\newblock Bag of tricks for efficient text classification.
\newblock {\em arXiv preprint arXiv:1607.01759}, 2016.

\bibitem{ding2019instruction}
Nan Ding and Samuel Williams.
\newblock {\em An instruction roofline model for gpus}.
\newblock IEEE, 2019.

\bibitem{wang2017malware}
Wei Wang, Ming Zhu, Xuewen Zeng, Xiaozhou Ye, and Yiqiang Sheng.
\newblock Malware traffic classification using convolutional neural network for
  representation learning.
\newblock In {\em 2017 International conference on information networking
  (ICOIN)}, pages 712--717. IEEE, 2017.

\end{thebibliography}

\end{document}